# A Projected Entropy Controller for Transition Matrix Calculations

David Yevick

Department of Physics

University of Waterloo

Waterloo, ON N2L 3G7

**Abstract:** We define the projected entropy $S(T)$ at a given temperature $T$ in the context of an Ising model transition matrix calculation as the entropy associated with the distribution of Markov-chain realizations in energy-magnetization, $E-H$, space. An even sampling of states is achieved by accumulating the results from multiple Markov chains while decrementing $1/T$ at a rate proportional to the inverse of the effective number, $\exp(S(T))$, of accessible projected states. Such a procedure is both highly accurate and far simpler to implement than a previously suggested method based on monitoring the evolution of the $E-H$ distribution at each temperature. [1] We further demonstrate a transition matrix procedure that instead ensures uniform sampling in physical entropy.

**PACS Codes:** 02.70.Rr, 02.70.Uu, 05.10.Ln, 02.50.Ng

**Keywords:** Statistical methods, Transition matrix, Ising model

**Introduction:** This paper considers the determination of the statistical behavior of one or more global variables $\vec{E}(\vec{\alpha})$ that depend on numerous stochastically varying local quantities $\vec{\alpha}$. While this information can be directly obtained by Monte-Carlo (randomly) sampling the parameter space, $\vec{\alpha}$, the computation time is often excessive if the properties of interest are associated with low probability regions of $\vec{E}(\vec{\alpha})$.

Accordingly, numerically efficient Markov-chain based procedures such as the multicanonical [2] [3] [4] [5] [6] [7] [8] [9] [10] [11] [12] and Wang-Landau [13] [14] [15] methods bias the local variables $\vec{\alpha}$ through an acceptance rule such that instead of the individual components of $\vec{\alpha}$, physically relevant ranges of the $\vec{E}(\vec{\alpha})$ are nearly uniformly sampled. Such methods can additionally be employed to construct a transition matrix $\mathbf{T}$ where $\mathbf{T}_{ij}$ corresponds to the probability that a Markov chain state in a histogram bin $E_i$ transitions to bin $E_j$ in a single unbiased Markov step $\delta\vec{\alpha}$ recorded before the application of the acceptance rule. [1] [16] [17] [18] [19] [20] [21] [22] [23] [24] [25] [26] The normalized eigenvector of $\mathbf{T}$ with unit eigenvalue, which can be obtained by repeatedly multiplying an initially random vector by $\mathbf{T}$ then coincides with the desired probability distribution $p(E)$. [27] The transition matrix method can also be accelerated by incorporating renormalization techniques that estimate the transition matrix associated with a system by iteratively convolving subsystem matrices. [28] [29] [30]

When applied to the Ising model, the transition matrix procedure generates a proposed configuration by flipping a single, randomly generated spin, increments the unnormalized transition matrix element corresponding to this unbiased transition by unity, and finally biases the sample space by applying an appropriate transition rule. Since the transition matrix procedure therefore incorporates all attempted transitions from a given state, simple algorithms based on e.g. transition probabilities between microscopic states [16] [17], the ratio of transition matrix elements [31] [32] [17], and the exclusion of transitions to bins that have been previously visited a larger number of times. [23] [27] [33] can be employed to generate $\mathbf{T}$. While the method is formally independent of the transition rule, the specific features of the rule indirectly affect the accuracy of computationally sensitive quantities such as the specific heat near the critical temperature. In fact, standard procedures based on accumulating transition matrix elements over a series of canonical, Metropolis [34] calculations at inverse temperatures that vary progressively from a large initial value to nearly zero display far greater precision than quasi-microcanonical methods that sample $\vec{E}(\vec{\alpha})$ within a narrow but similarly shifting region. [35] [36] [29] The accuracy can however generally be enhanced by increasing the number of realizations for temperatures near the critical temperature and by accumulating transitions from multiple independent Markov chains.

The influence of the transition rule on the transmission matrix was recently heuristically explained by referring to the observed nearly constant diffusion velocity of the Markov chain in the two-dimensional $E-H$ space. [29] For a thermal Metropolis transition rule, near the critical temperature, $T_c$, a far larger area of this plane is physically accessible to the Markov chain, which accordingly requires a greater number of steps to diffuse and reach a stable distribution. Accordingly, the number of realizations at each temperature can be determined by monitoring the convergence of the normalized histogram of samples as a function of magnetization, yielding a procedure that dynamically adapts to variations in the size of the accessible projected phase space. [1] Observe that a somewhat more accurate but structurally complex technique would be to enforce convergence over the entire two-dimensional $E-H$ distribution at each temperature.

This paper presents a more convenient procedure for controlling automatically the thermal evolution of a transition matrix computation based on the *projected entropy* of the state space onto the $E-H$ plane. In the $k$:th temperature step of the calculation, for which $T=T_k$, the transition matrix elements are updated with the unbiased transitions obtained as each of a set of $N_{\text{chains}}$ Markov chains is evolved $N_{\text{steps}}$ steps subject to the Metropolis acceptance rule. At the same time, a histogram, $\Omega_{\text{realizations}}(E_i, H_j, T_k)$ is recorded that accumulates the total number of realizations at $T_k$ that possess energies and magnetizations equal to $E_i$ and $H_j$. The projected entropy

$$S(T_k) = -\sum_{i,j} p(E_i, H_j, T_k) \log p(E_i, H_j, T_k) \tag{1}$$

is subsequently evaluated, where the probability distribution of a projected state in the $E-H$ plane, $p(E_i, H_j, T_k) = \Omega_{\text{realizations}}(E_i, H_j, T_k) / \sum_{i,j} \Omega_{\text{realizations}}(E_i, H_j, T_k)$. The exponential of the projected entropy then yields the effective number, $\Omega_{\text{states}}(T_k) = \exp(S(T_k))$, of $E-H$ states that are visited at the specified temperature. Hence to sample uniformly the $E-H$ space the inverse temperature is progressively decreased such that the $k$:th inverse temperature step is given by

$$\left(\Delta\left(\frac{1}{T}\right)\right)_k = -\min\left(\Delta_{max}, \frac{\chi}{\Omega_{realizations}(T_{k-1})}\right) \qquad (2)$$

where the calculations of this paper proceed from an initial large inverse temperature $1/T_0$ to a small final value $1/T_f$. The constant $\chi$ is empirically determined as described below to ensure that the inverse temperature step is sufficiently small near the critical temperature.

An alternative to the above technique based on the exact rather than the projected entropy distribution can be constructed from the partial results for the transition matrix at each intermediate temperature, $T_i$, as the inverse temperature is raised from a near zero value in a stepwise fashion. Here the detailed balance condition

$$\log \Omega^{(i)}(E_i) + \log \mathbf{T}_{ii-1} = \log \Omega^{(i-1)}(E_{i-1}) + \log \mathbf{T}_{i-1i} \qquad (3)$$

is employed as in [37] recursively to determine the density of states (to within a constant factor). However in this paper, to our knowledge for the first time, we generalize this procedure by employing at the *i*:th inverse temperature step only the contributions to the transition matrix elements obtained at the previous calculation steps. The resulting estimate of the density of states extends (at low energies) to the smallest energy for which the amplitude of the $T_{i-1}$ Boltzmann distribution is substantial and hence $\mathbf{T}_{ii-1}$ and $\mathbf{T}_{i-1i}$ are both non-zero. The corresponding canonical entropy as a function of energy, $\log \Omega^{(i)}(E)$, is then obtained from $S(T_i^{-1}) \propto \sum_j e^{-\beta E_j}\left(\log e^{-\beta E_j} - \log Z(T_i^{-1})\right)\Omega^{(i)}(E_j)$, where $Z(T^{-1})$ denotes the partition function. Then setting the succeeding inverse temperature step equal to

$$\left(\Delta\left(\frac{1}{T}\right)\right)_k = \frac{\chi'}{\frac{d \log S(T_k^{-1})}{dT^{-1}}} \approx \frac{\chi''}{S(T_k^{-1}) - S(T_{k-1}^{-1})} \qquad (4)$$

ensures that the steps are evenly spaced in entropy. Accordingly, $S_N = S_0 + N\Delta S$ or $\Omega_N = \Omega_0 (\exp(\Delta S))^N$, which coincides with the objective of e.g. nested sampling. [38] [39] Unlike this procedure, however, the present technique does not require numerous Markov steps to insure that the added Markov chains achieve statistical independence with the surviving chains.

**Numerical results:** To demonstrate the accuracy of the projected entropy controller the standard benchmark calculation of the specific heat of the two dimensional Ising model with zero external magnetic field, periodic boundary conditions, and a unit amplitude ferromagnetic interaction will be performed for an array of $32 \times 32$ integral spins. As evident from the figures in e.g. Ref. [29], the effective number of accessible states in the $E - H$ plane exhibits a pronounced maximum at a temperature slightly above the critical temperature, $T_c = 1/1.743$, in normalized units. Accordingly, the number of steps required for a single Markov chain to visit the entire accessible projected region and thus to populate sufficiently all relevant transition matrix elements increases markedly near the critical temperature.

The suitability of the effective number of projected states (the exponential of the projected entropy of Eq. (1)) as a metric is first clarified by examining its behavior as a function of the number of the Markov chain realizations at 12 evenly spaced values of the normalized inverse temperature between 1.68 and 1.73 that are slightly below the inverse critical temperature. Starting in a fully magnetized zero entropy configuration at each temperature yields the result of Fig. 1 for $N_{steps} = 2 \times 10^9$ and $N_{chains} = 1$. The associated number of realizations at each temperature as a function of the magnetization, scaled such that the maximum of each curve is unity is presented in Fig. 2.

Note that while the entropy converges least rapidly at $T \approx T_c$, the projected state distribution most nearly resembles a constant function of magnetization at somewhat higher temperatures. This implies that the slow convergence rate near the critical temperature is associated with the low frequency of displacements from positive to negative magnetization regions, which affects the population of the transition matrix elements in the region of small $H$. Somewhat above the critical temperature, however, this frequency is greatly enhanced, resulting in both a distribution that for a restricted range of temperatures is nearly independent of $H$ and a more rapidly converging projected entropy.

Widely separated regions of configuration space can be simultaneously sampled by employing numerous mutually independent Markov chains. In vector-oriented languages such as MATLAB, such a procedure is considerably more efficient than employing a large number of steps to insure the diffusion of a single Markov chain over the physically accessible $H - E$ states. To illustrate, Figs. 3 and 4, correspond to the calculations of Figs. 1 and 2, respectively, but with $N_{chains} = 100$ and $N_{steps} = 4 \times 10^6$. While a direct comparison between the two sets of figures is misleading since each horizontal axis unit in Figs. 1 and 2 corresponds requires 100 times more realizations than in Figs 3 and 4, the greater accuracy of the latter figures is manifest.

To employ the projected entropy controller of Eq. (2), the computational parameter $\chi$ must first be specified. Since the number of temperature steps and hence the computation time decreases monotonically with $\chi$, the smallest possible value of $\chi$ consistent with the available CPU time should be employed. To illustrate, Fig. 5 displays the evolution of $1/T_k$ with the temperature iteration number, $k$, for a maximum step length of $\Delta_{max} = 0.1$ and $\chi$ equal to $2.5 \times 10^3$ (dashed line) and $1.0 \times 10^3$ (solid line). As the latter value significantly enhances the number of temperature evaluations slightly above the transition temperature where the convergence is slowest, it is utilized in the specific heat calculation below. An critical feature of the method is that the curves of Fig. 3 are effectively independent of $N_{steps}$, thus for example, nearly 120 temperature steps will always be required to perform a transition matrix calculation for $\chi = 10^3$.

A direct comparison of the projected entropy method with the reference calculation for the adaptive procedure of Ref. [1] can be achieved by setting $N_{chains} = 1$ and $N_{steps} = 1.3 \times 10^7$ in the projected entropy technique. Repeating this calculation 36 times and forming a histogram from the resulting specific heat curve maxima yielded the result of Fig. 6, which compares with the exact result of 1.9045. [40] [37] The evaluation of a specific heat curve required 212 minutes on an Intel i7-3770K processor. In all calculations the initial inverse temperature was set to $1/T_0 = 3.2$ and the calculation then performed with decreasing inverse temperatures following the prescription of Eq. (2) until a final inverse temperature of $1/T_f = 0.1$ was attained according to the schedule of Fig. 5. The initial spin state for the Markov chain at $T_k$ was provided by the final realization at temperature $T_{k-1}$ ensuring a more rapid convergence. Fig. 7 displays a histogram of the total number of realizations employed in the calculations, which compares to $\approx 1.67 \times 10^9$ in Fig. 6 of Ref. [1]. While the accuracy is slightly less than the procedure of the reference, this is compensated by the greater simplicity of the algorithm.

In e.g. MATLAB, the efficiency of the calculation can be greatly increased by employing multiple chains. In Figs. 8 and 9, analogous results to those of Figs. 6 and 7 are displayed with $N_{chains} = 10$ and $N_{steps} = 1.3 \times 10^6$. In this calculation, MATLAB only required 6.4 minutes to generate a specific heat curve, which included data from 10 chains. A direct comparison between computation times is however misleading since 6 computations were in all cases performed simultaneously on an 8 logical processor CPU and were therefore affected by memory limitations and other running processes. Finally Fig. 10, for which $N_{chains} = 100$, $N_{steps} = 4 \times 10^6$, the total number of realizations is $5 \times 10^{10}$ and the CPU time 193 minutes for a specific heat curve evaluation, clearly demonstrates the potential accuracy of the method.

In order to generate from Eq. (4) a temperature schedule similar to that of Fig. (5), the intermediate results for the density of states, $\Omega^{(i)}(T_i)$, are evaluated from the transition matrix elements accumulated during the calculations at inverse temperatures $4.5, 4.4 \ldots, 1/T_{i-1}$. The curves for all $\Omega^{(i)}(T_i)$ are overlaid in Fig. 11 and the terminating position of each curve is distinguished by a row of markers. Clearly, each partial result closely approximates the final density of states curve in overlapping regions in analogy to [41] (it should be also be noted that this technique can also be applied to determine the density of states up to a normalizing constant within any limited region of energy). The associated canonical entropy at each inverse temperature is calculated and displayed in Fig. 11 where the dashed line indicates the critical temperature. Inserting the finite difference of the entropy into Eq. (4) finally yields Fig. 12 which, as expected, implies that a far smaller inverse temperature steps should be taken in the vicinity of the critical temperature. The discrepancy between this prescription and Fig. 5 is at least partially associated with the intrinsic error of the backward finite difference expression for the entropy derivative.

**Discussion and Conclusions:** That the accuracy of the transition matrix can be monitored by projecting the full microscopic state space of the Markov chain realizations onto a space defined by the variables that are directly related to the transition matrix could have numerous, foundational implications. For example, a refined version of the controller that would include additional variables into the expression for the projected entropy might be applicable to more complex systems. Another method derives from the observation that, as mentioned above, the evolution of the full $E-H$ state distribution rather than just the distribution of states as a function of magnetization as in Ref. [1] provides an alternative metric to establish if a computation has properly converged at a given temperature. By analogy, the computed effective number of projected states together with the diffusion velocity of the Markov chain could similarly be employed to determine the number of realizations required for convergence at each temperature rather than the magnitude of the temperature step. Alternatively, if the convergence of the entropy is monitored at each temperature, the number of Markov steps can be adjusted dynamically.

While such procedures could presumably offer additional accuracy, the projected entropy controller of this paper provides a direct and simple technique for controlling the temperature schedule in transition matrix procedures. Further, the accuracy can be adjusted through a single parameter that is almost independent on the number of realizations employed at each temperature step. This provides an intrinsic advantage over, for example, our earlier adaptive transition matrix procedure, [1] that contain multiple parameters that are not easily optimized. As well, in vector programming languages, greatly increased computational efficiency can be achieved with minimal additional programming effort by introducing multiple simultaneous Markov chains.

**Acknowledgements:** The Natural Sciences and Engineering Research Council of Canada (NSERC) is acknowledged for financial support.

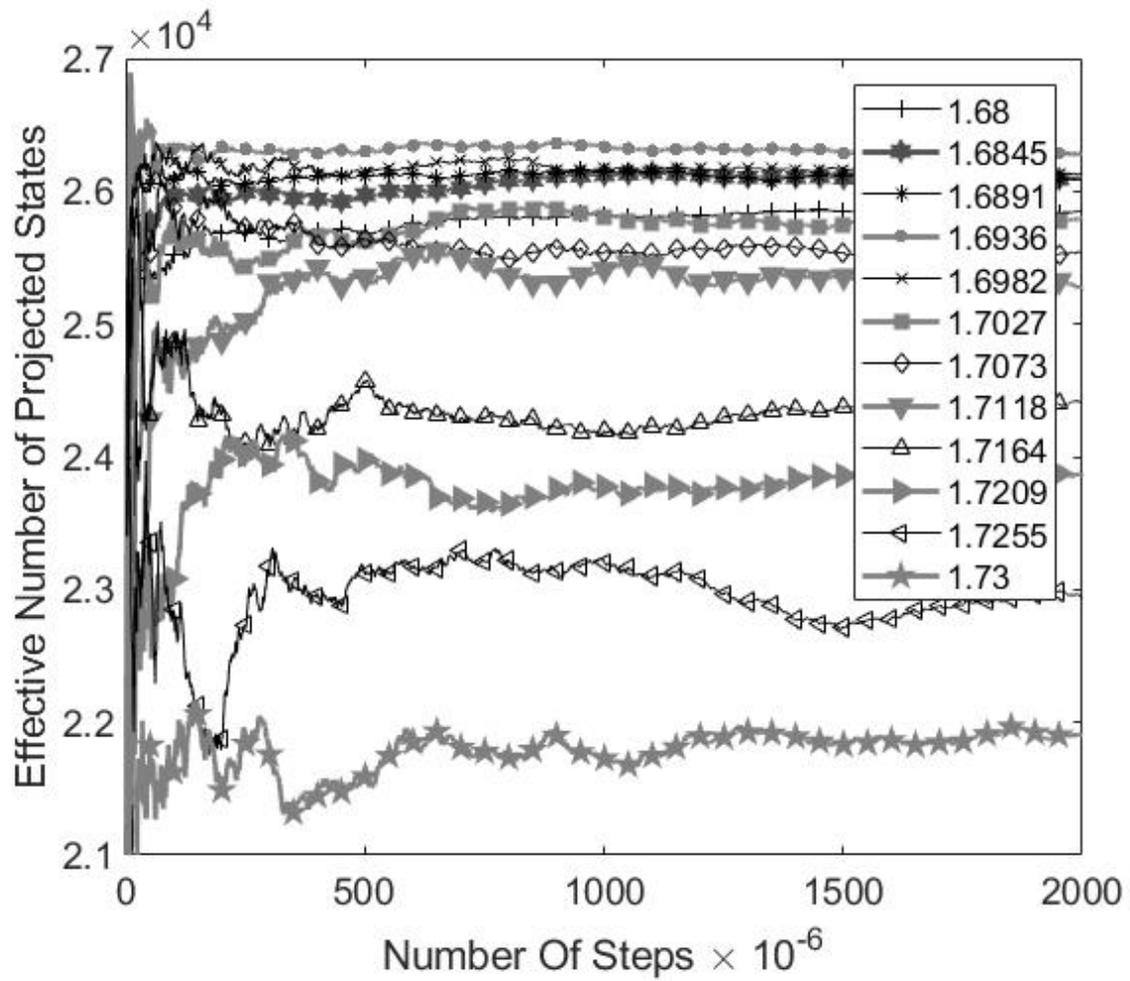

Figure 1: The effective number of projected states as a function of the number of the Markov chain realizations for a single Markov chain and the normalized inverse temperatures indicated in the legend.

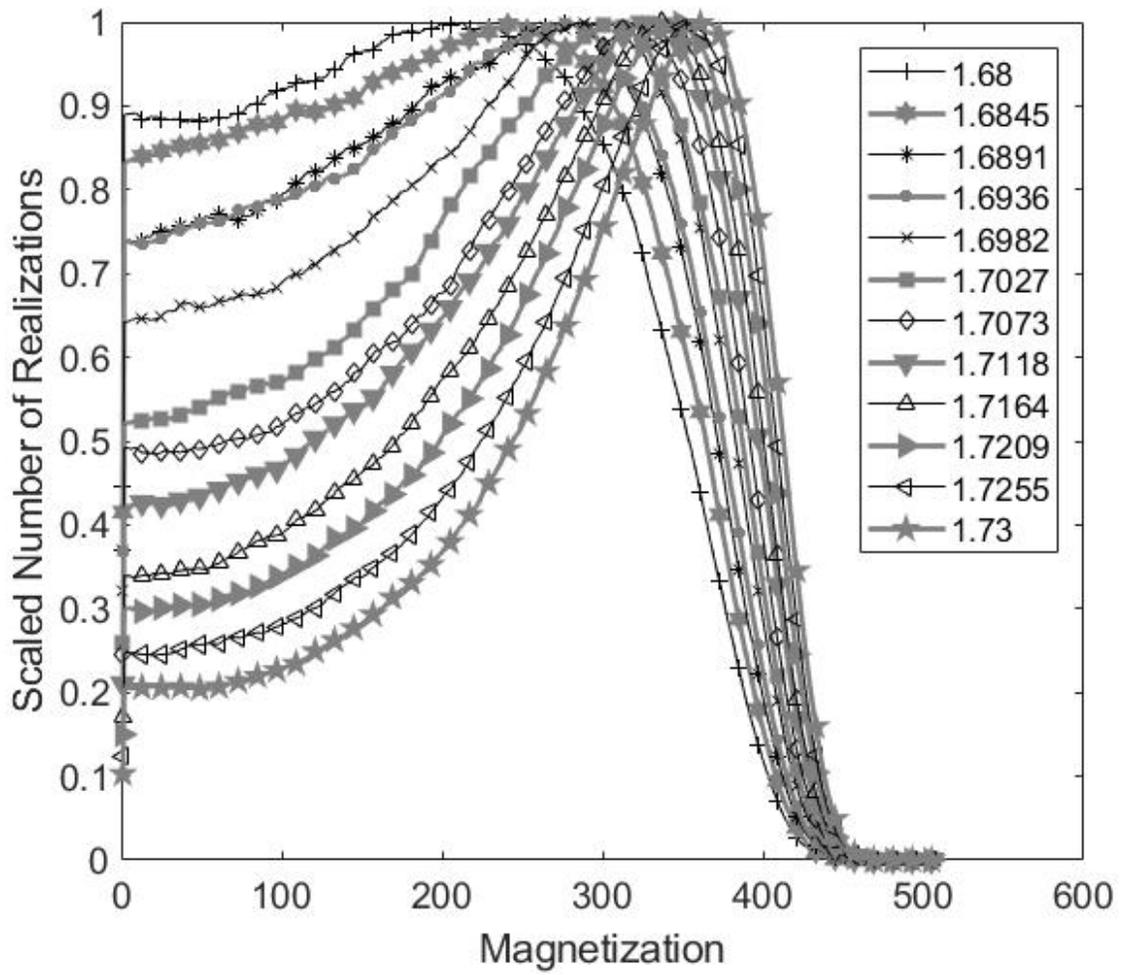

*Figure 2: The distribution of visits of the Markov chain at each temperature in Fig. 1 as a function of the magnetization, scaled such that the maximum of each curve is unity.*

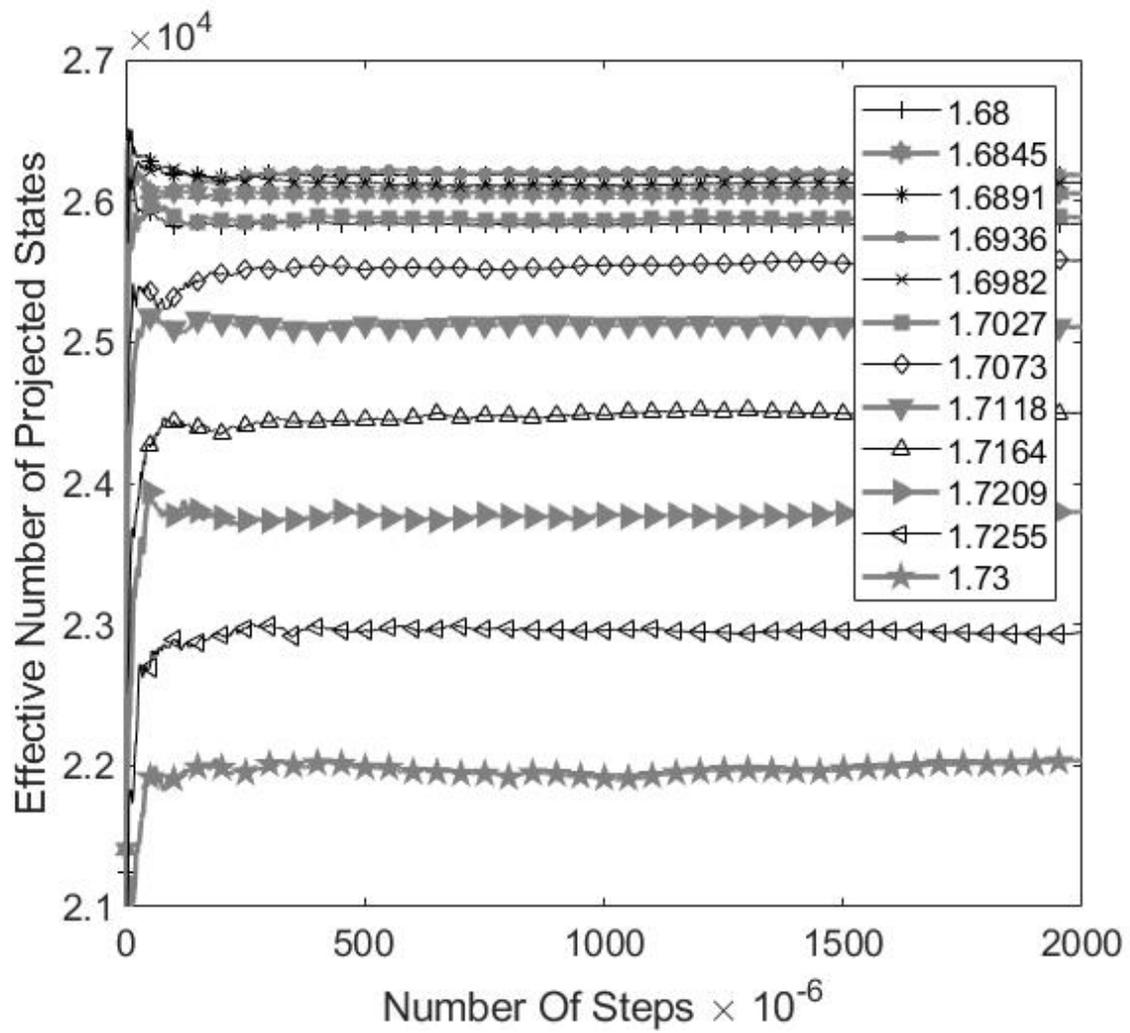

Figure 3: As in Fig. 1 but for 100 simultaneous Markov chains.

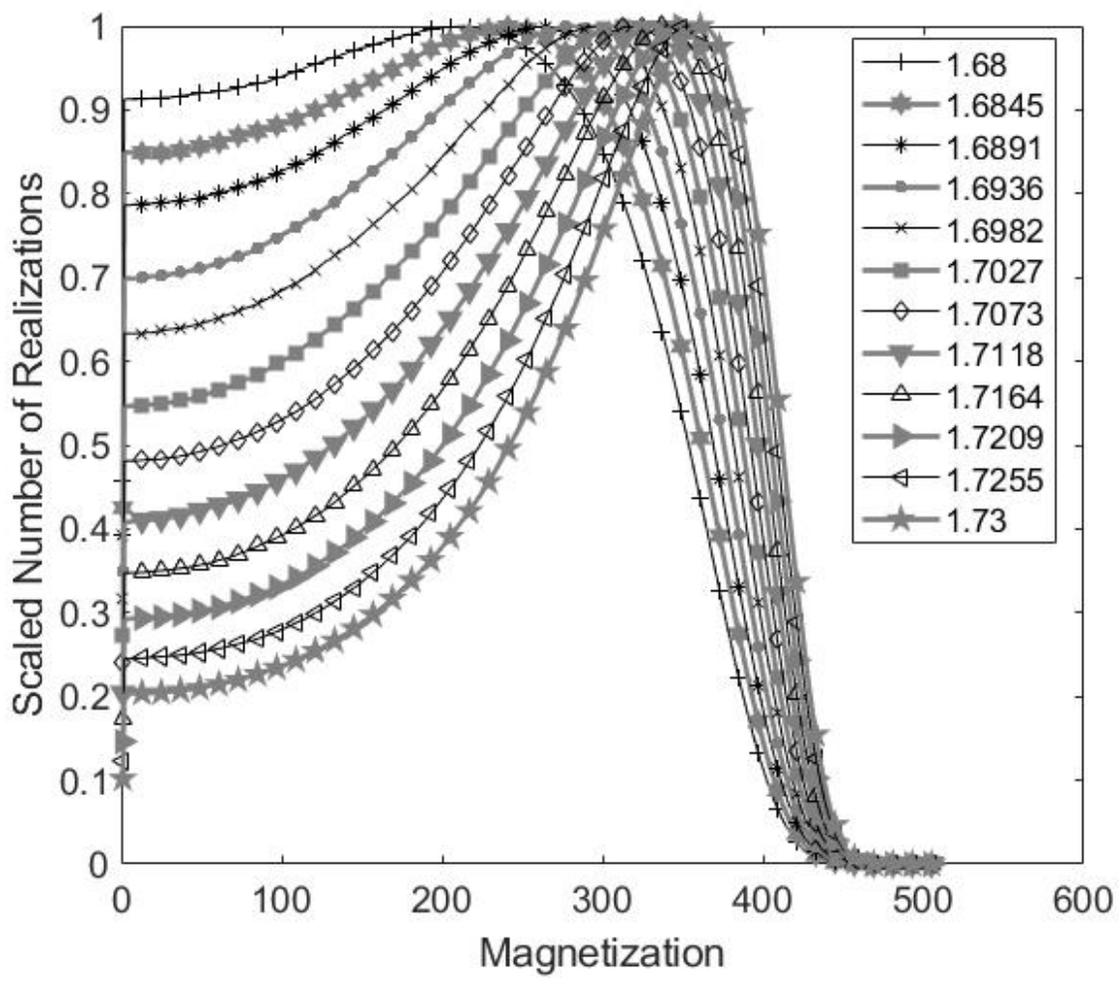

*Figure 4: As in Fig. 2 but for 100 simultaneous Markov chains*

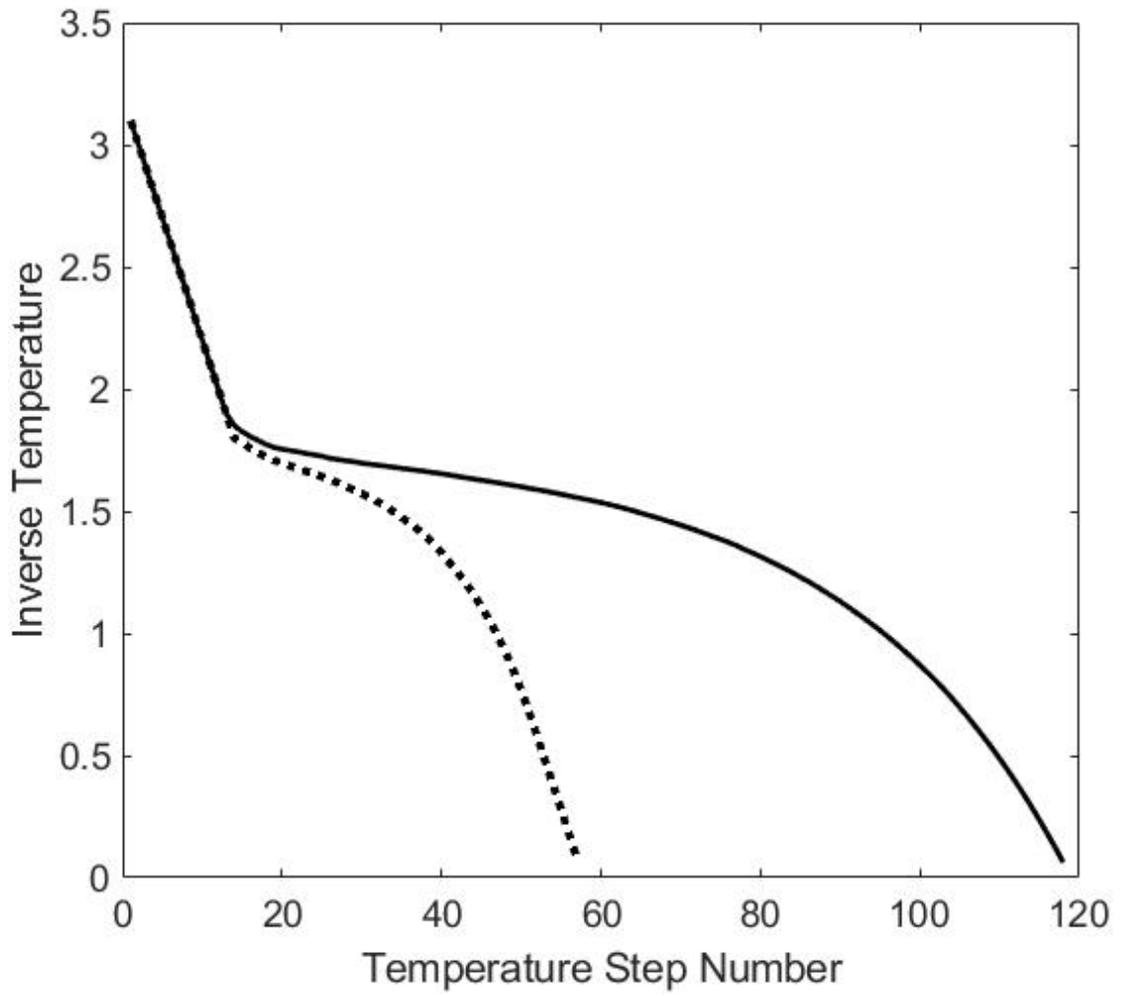

Figure 5:The inverse temperature as a function of iteration steps for a maximum step length of $1/T = 0.1$ and the computational parameter $\chi$ equal to $2.5 \times 10^3$ (dashed line) and $1.0 \times 10^3$ (solid line).

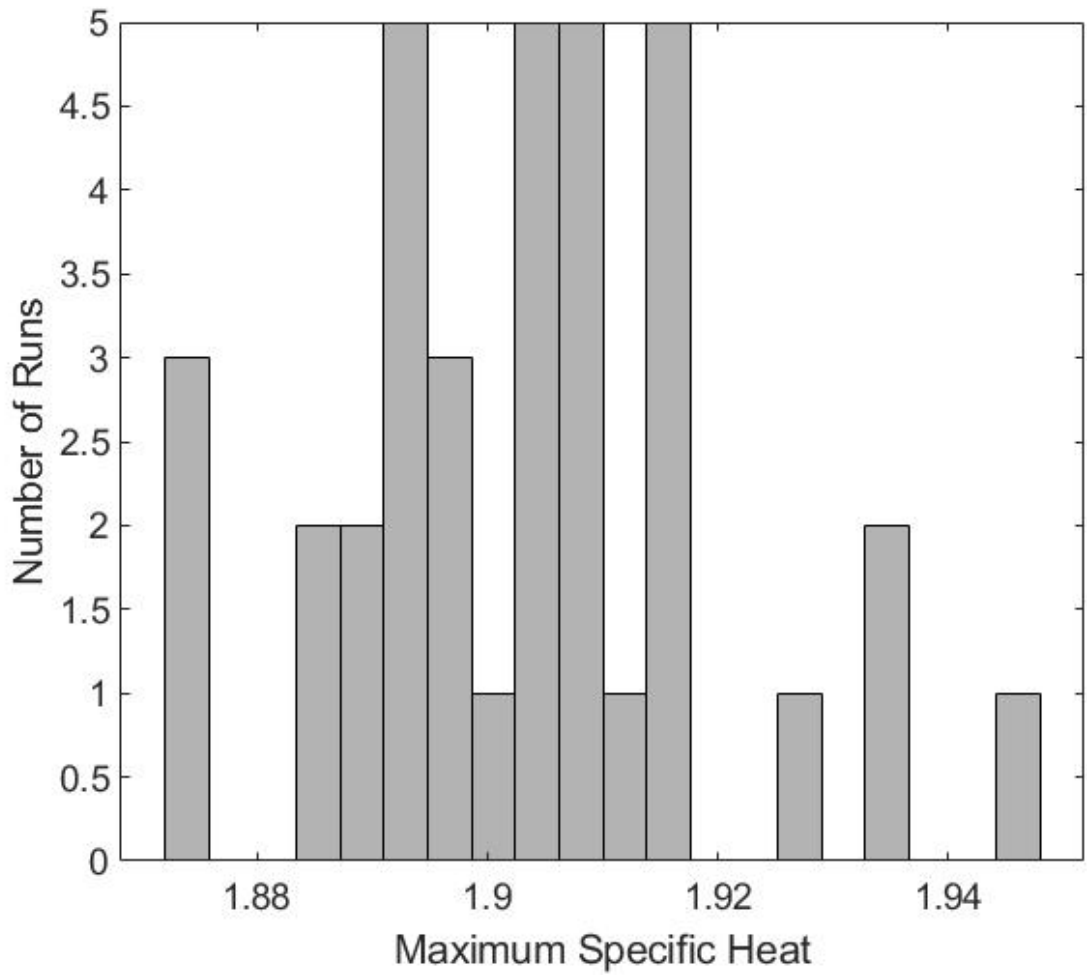

Figure 6: The histogram of the maxima of the specific heat curves obtained after 36 independent calculations each of which employed a single Markov chain with $1.3\times10^7$ realizations.

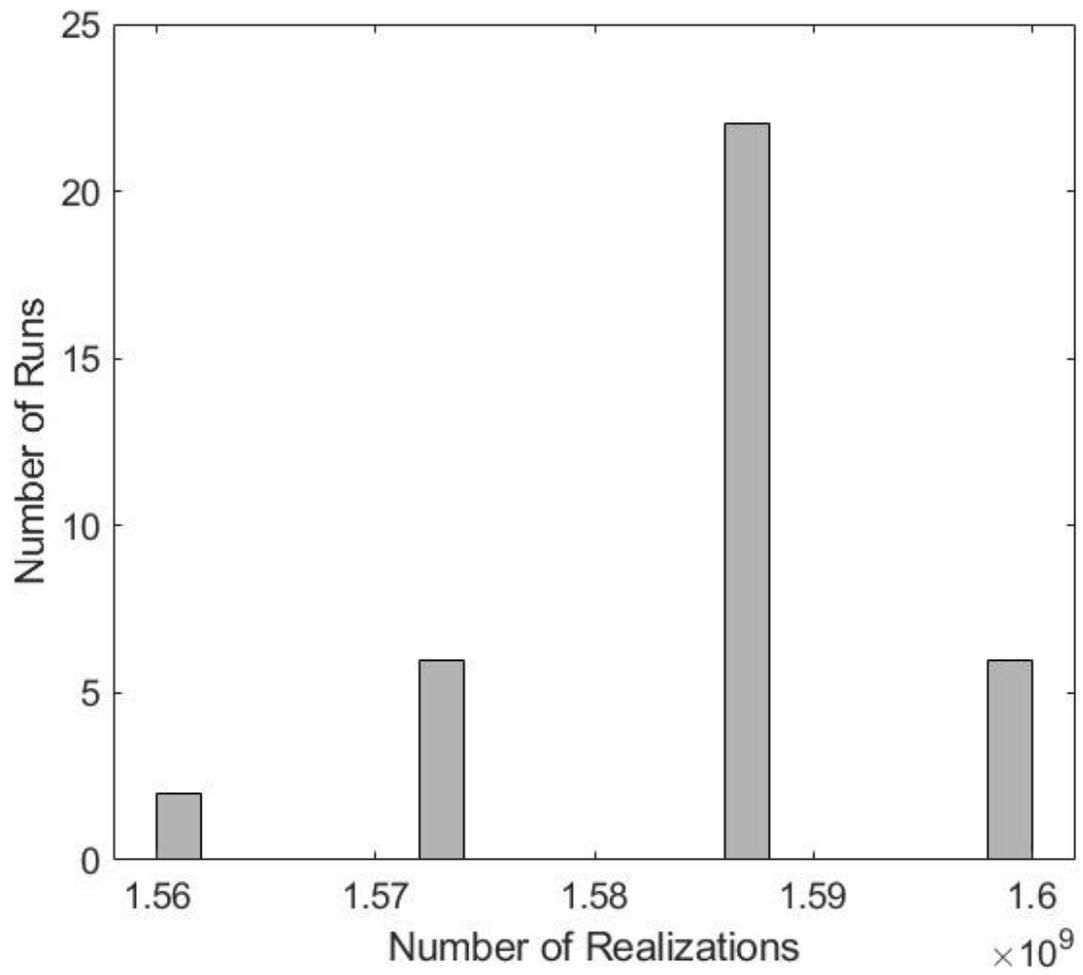

Figure 7: A histogram of the total number of realizations employed in each of the 36 calculations of Fig. 6

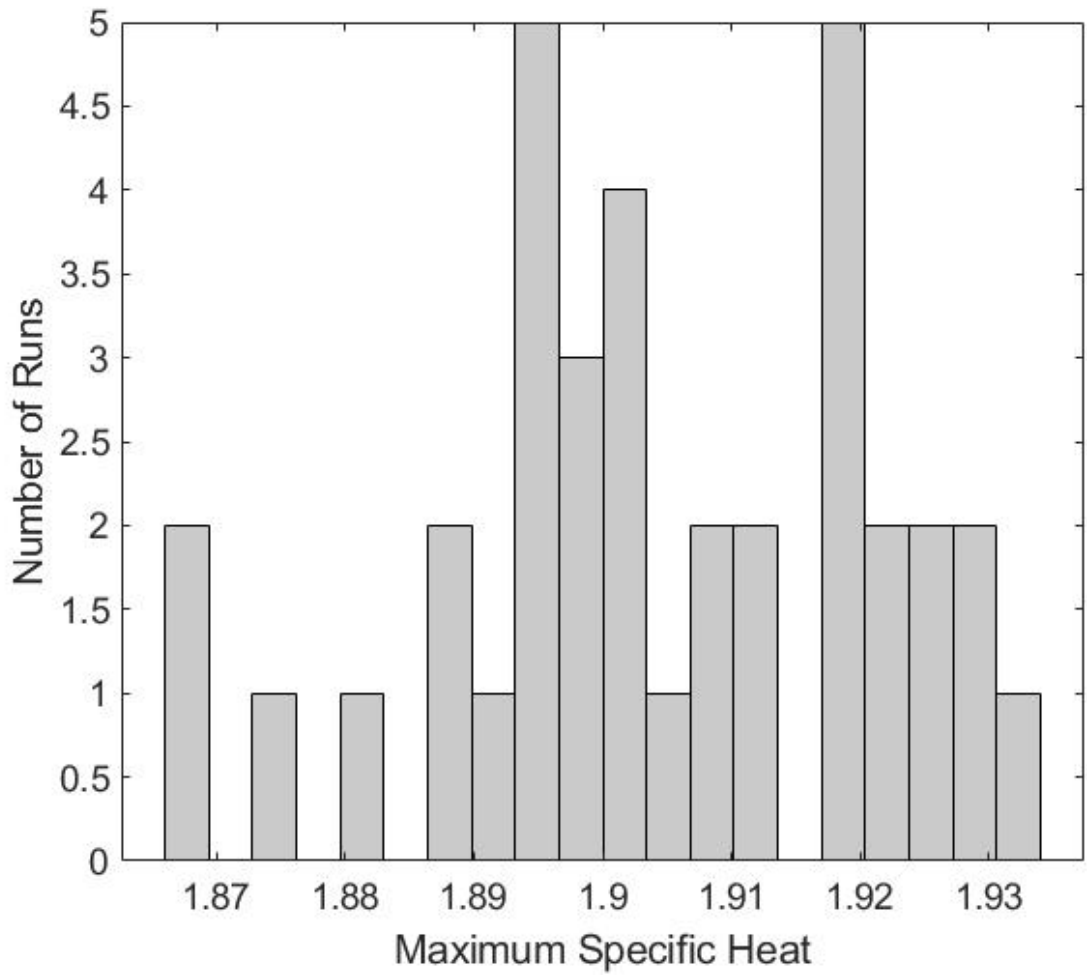

Figure 8: As in Fig. 6 but with 10 Markov chains with $1.3 \times 10^6$ realizations per chain

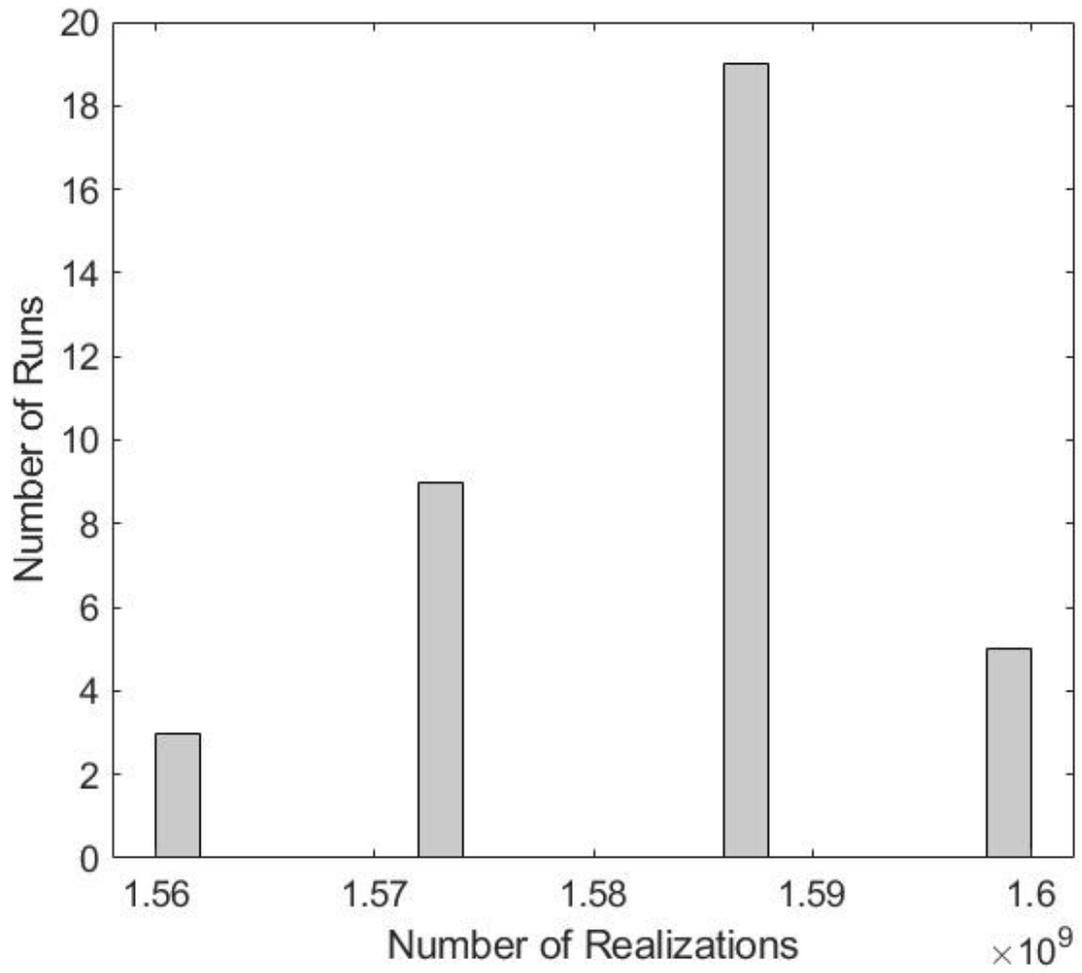

Figure 9: A histogram of the total number of realizations employed in each of the 36 calculations of Fig. 8.

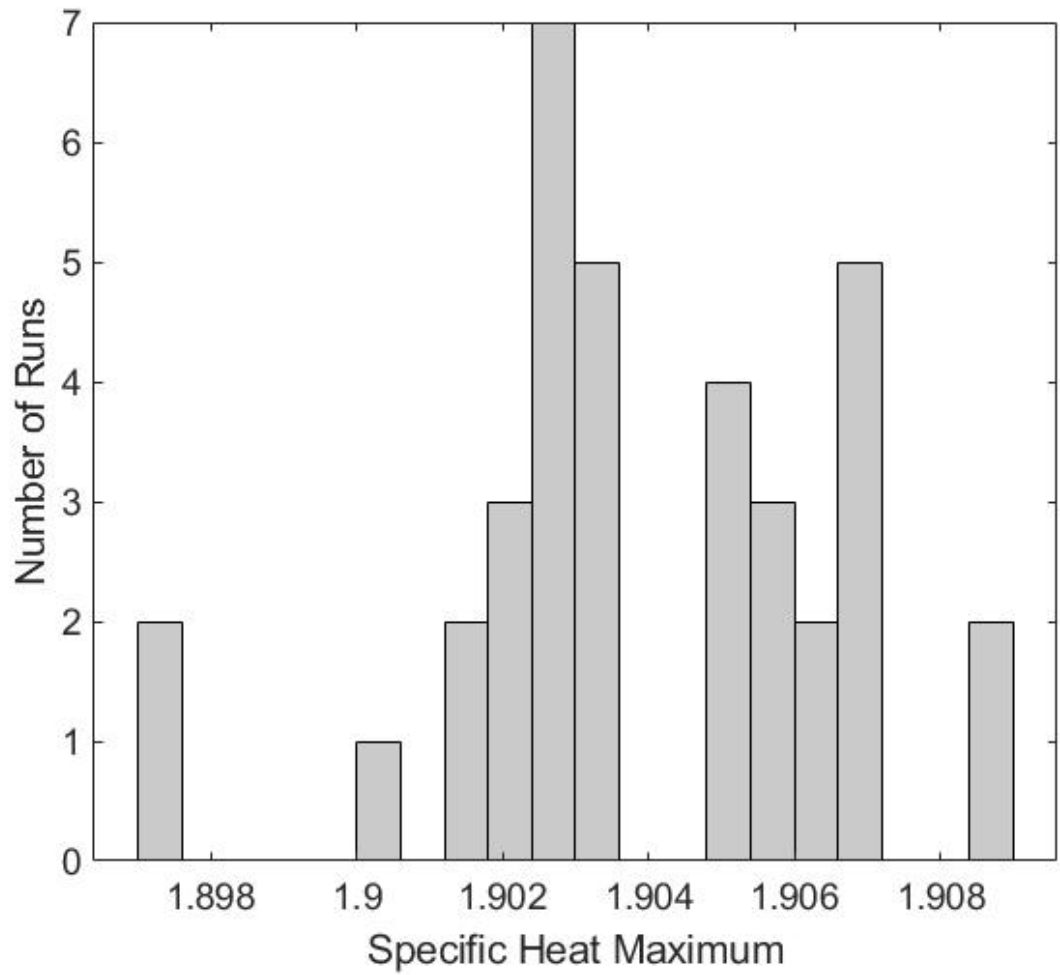

Figure 10: The histogram of the maxima of the specific heat curves obtained after 36 independent calculations each of which comprised a hundred $4\times10^6$ realization Markov chains.

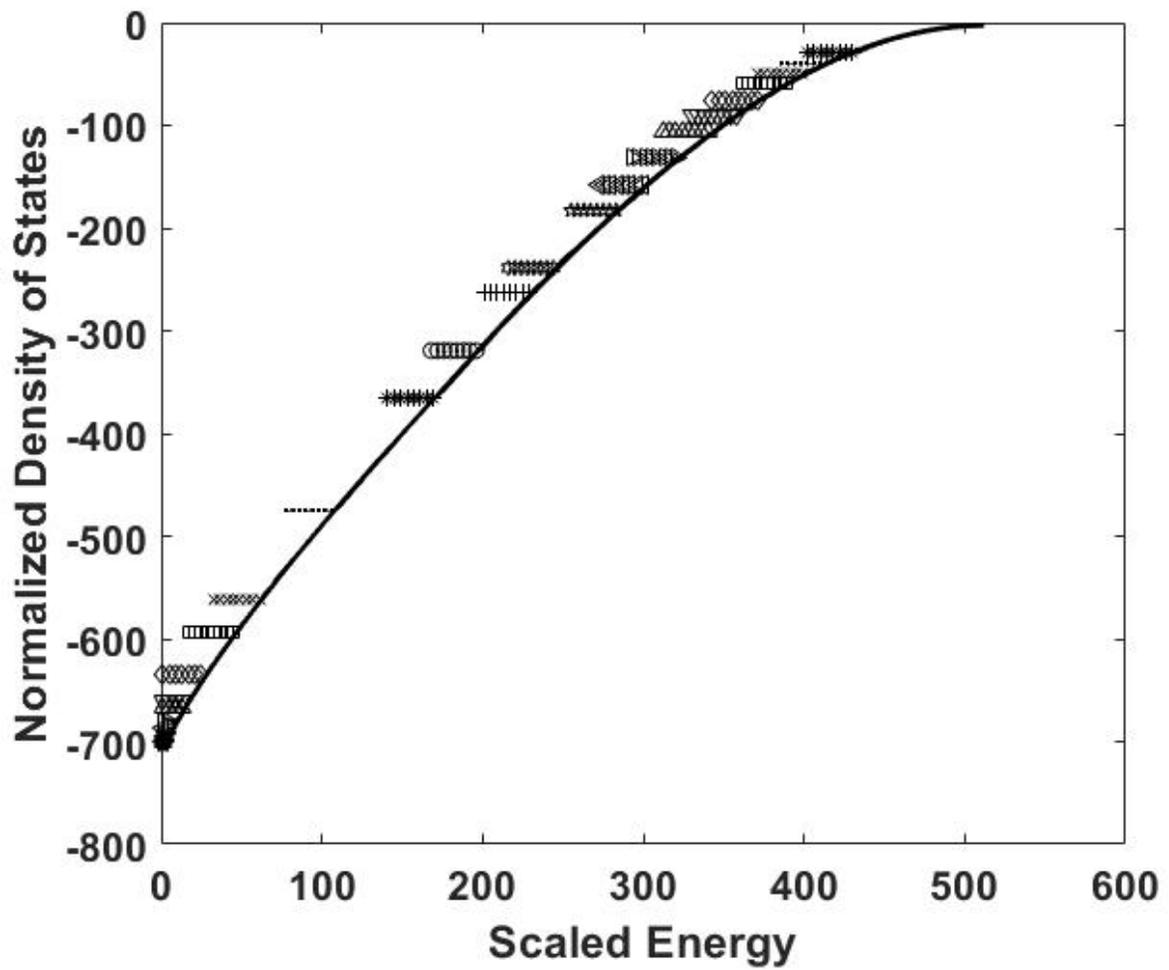

Figure 11: Intermediate results for the i:th partial density of states evaluated from transition matrix elements accumulated at inverse temperatures $4.5, 4.4\ldots, 1/T_{i-1}$. The markers designate the endpoints of each curve.

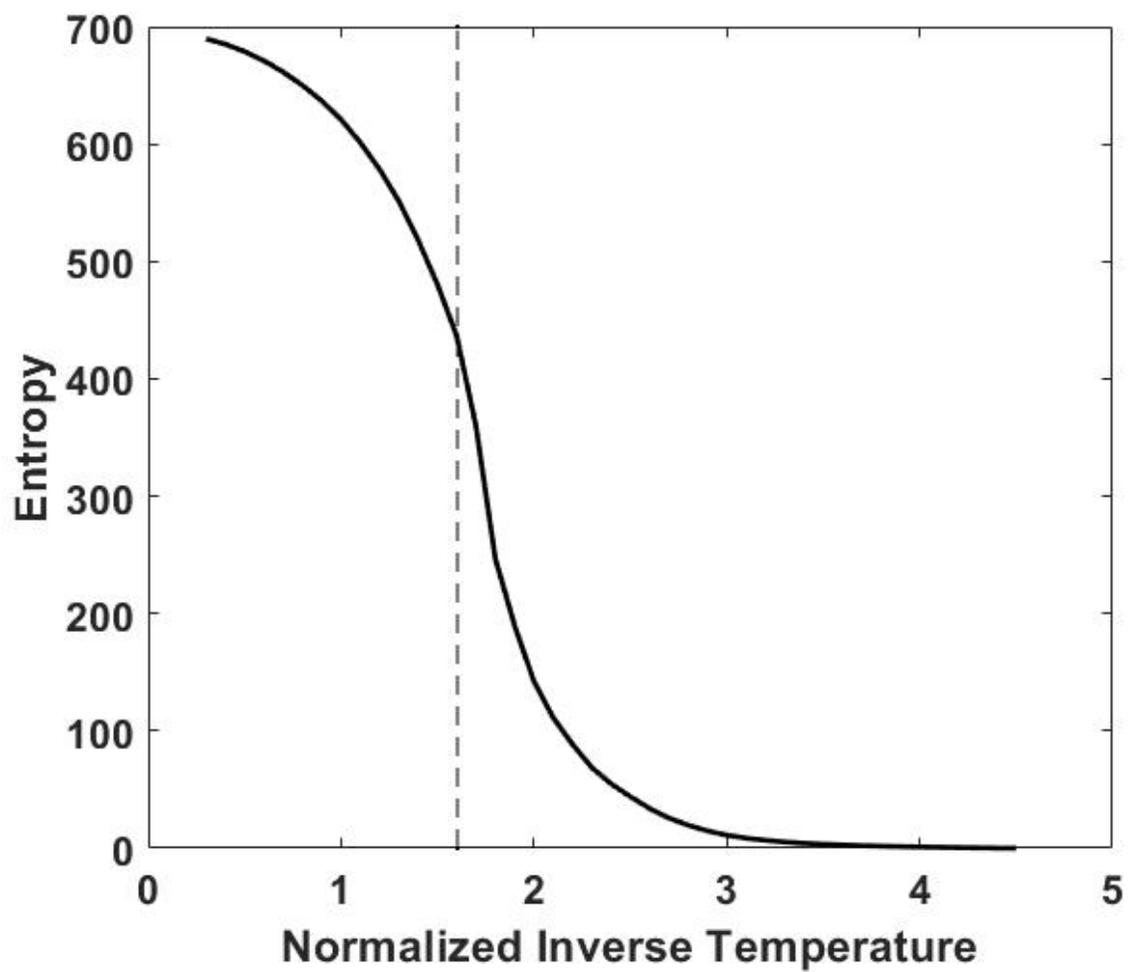

*Figure 12: The entropy as a function of normalized inverse temperature, where the dashed line indicates the critical temperature*

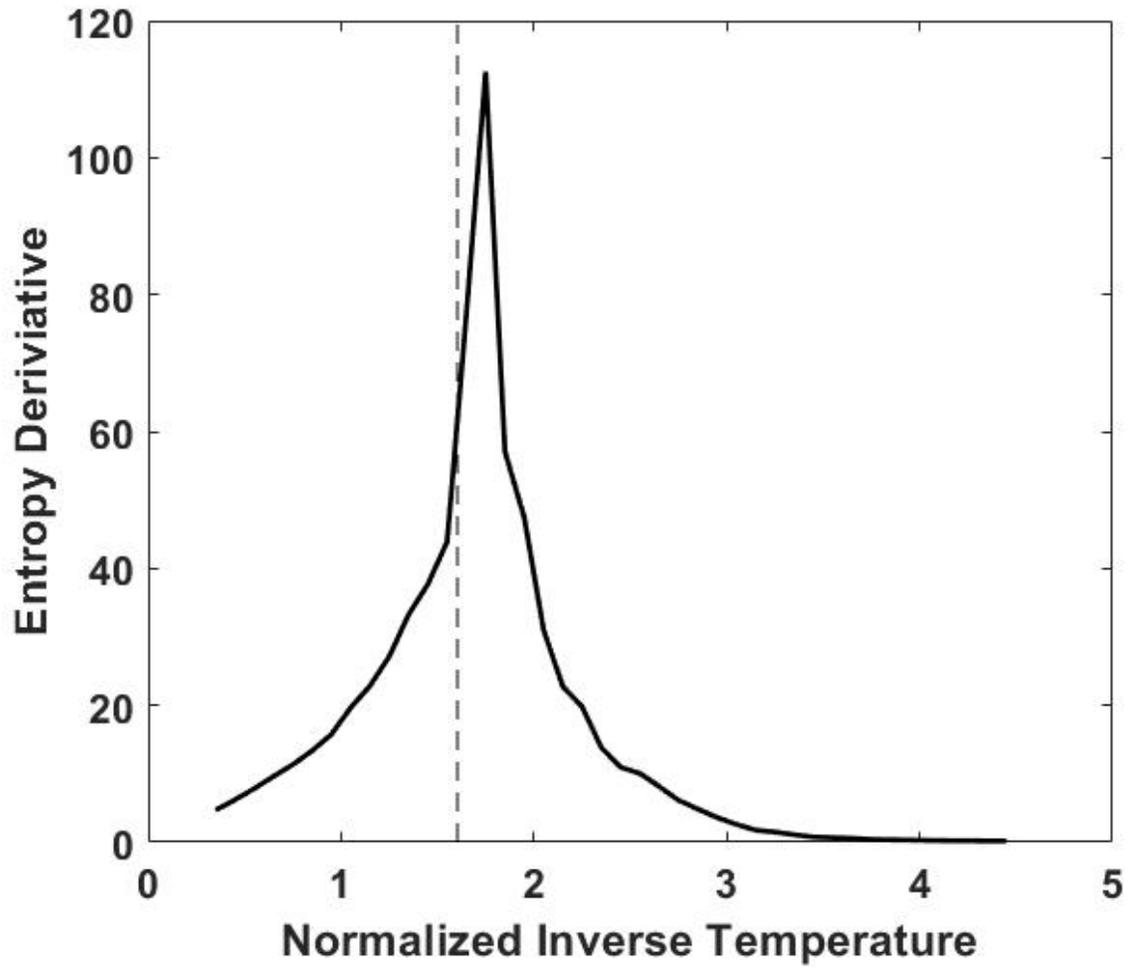

Figure 13: The finite difference of the entropy of the previous figure, proportional to the inverse temperature step.